\documentclass[%
 reprint,
 amsmath,amssymb,
 aps,
]{revtex4-1}
\pdfoutput=1
\usepackage{graphicx}
\usepackage{amsmath}
\usepackage{amssymb}
\usepackage{dsfont}
\usepackage{tensor}
\usepackage[font=footnotesize,labelfont=bf]{caption}

\begin{document}
\pdfoutput=1
\preprint{APS/123-QED}

\title{Microscopic modeling of the effect of phonons on the optical properties of solid-state emitters}

\author{{ Ariel Norambuena$^{1,2}$, Sebasti\'an A. Reyes$^{1,2}$, Jos\'e. Mej\'ia-Lop\'ez$^{1,2}$, Adam Gali $^{3,4}$ and Jer\'onimo. R. Maze$^{1,2}$}\\
\vspace{0.5 cm}
{\small \em $^1 $Faculty of Physics, Pontificia Universidad Cat\'olica de Chile, Avda. Vicu\~{n}a Mackenna 4860, Santiago, Chile.\\
$^2$Center for Nanotecnology and Advanced Materials CIEN-UC, Pontificia Universidad Cat\'olica de Chile, Avda. Vicu\~{n}a Mackenna 4860, Santiago, Chile.\\
$^3$ Department of Atomic Physics, Budapest University of Technology and Economics, Budafoki \'ut 8., H-1111 Budapest, Hungary.\\
$^4$Institute for Solid State Physics, Wigner Research Centre for Physics, Hungarian Academy of Sciences, P.O. Box 49, H-1525, Budapest, Hungary.}}
\date{\today}

\begin{abstract}
Understanding the effect of vibrations in optically active nano systems is crucial for successfully implementing applications in molecular-based electro-optical devices, quantum information communications, single photon sources, and fluorescent markers for biological measurements. Here, we present a first-principles microscopic description of the role of phonons on the isotopic shift presented in the optical emission spectrum associated to the negatively charged silicon-vacancy color center in diamond. We use the spin-boson model and estimate the electron-phonon interactions using a symmetrized molecular description of the electronic states and a force-constant model to describe molecular vibrations. Group theoretical arguments and dynamical symmetry breaking are presented in order to explain the optical properties of the zero-phonon line and the isotopic shift of the phonon sideband. 

\begin{description}
\item[PACS numbers]
78.67.Bf, 63.20.kp, 61.72.jn
\end{description}
\end{abstract}

\pacs{Valid PACS appear here}
\maketitle


\section{Introduction}
Vibrations play a crucial role in nano systems by modifying their optical line shape, preventing them from being described as simple two-level system \cite{1}. Several works have addressed the electron-phonon coupling to model the effect of vibrations on the optical properties of molecules \cite{2}, point defects \cite{3} and inter-band optical transitions in solids \cite{4}. This interaction is characterized, in most cases phenomenologically, by a spectral density function \cite{5,6,7} that is used to describe the dissipation dynamics due to acoustic phonons in a two-level system \cite{5}, the absorption \cite{8} and low temperature effects on the zero-phonon line transition \cite{6} in quantum dots 
that are strongly coupled to localized vibrations. There are few works that treat the electron-phonon interaction with microscopic models \cite{9}. The latter approach is particularly accurate for atomistic systems and highly demanded nowadays as researchers are able to engineer nanoscale devices where effectively few atoms are involved \cite{10}. Therefore, a deep understanding of this interaction is needed for controlling and engineering the optical properties of such systems. 

\vspace{0.3 cm}

Here we consider a microscopic model to study the electron-phonon interaction between the electronic states of a single negatively charged silicon-vacancy (SiV$^-$) center in diamond and lattice vibrations. We focus on the effect of phonons on the optical properties, i.e., the zero-phonon line (ZPL) transition and the phonon sideband associated to the emission or photoluminescence spectrum. On Section II we introduce to the electronic states of the SiV$^-$ center for which the optical emission will be calculated. Section III describes the vibrational degrees of freedom of a finite size lattice and the electron-phonon interaction between vibrations and the electronic states. Section IV introduces 
the model used to calculate the emission spectrum taking into account the symmetries of the electronic wavefunctions and vibrations. In particular, the spectral density function 
and its relation to the emission spectrum is introduced. Section V discusses the role of symmetry on the defect and finally Section VI takes into account these considerations 
to write the spectral density function for the SiV$^-$ center. 

\section{Negatively charged silicon-vacancy center in diamond}
In this section we present the bare ground and excited states from which the optical transitions will take place. 
The SiV$^{-}$ center is a point defect composed of six carbon atoms and an interstitial silicon atom. The symmetry group associated to this defect is the $C_{3v+i}$ group, a subgroup of the host crystal symmetry group $T_d$ \cite{16,20} (an equivalent group is $D_3$). In particular, the inversion symmetry with respect to the silicon atom leads to irreducible representations (IR) of the $C_{3v+i}$ group to be labeled by parity: $A_{1g}, A_{2g}, E_g$ ($g =$ gerade or even) and  $A_{1u}, A_{2u}, E_u$ ($u =$ ungerade or odd) representations \cite{20}. The electronic structure of this defect can be represented by one-electron hole system with electronic spin $S = 1/2$. In the absence of external perturbations the relevant electronic wavefunctions associated to the electron hole representation are 
\begin{eqnarray}
|\Psi_{gx,gy}^{(0)} \rangle &=& e_{gx,gy}^{\mbox{\scriptsize  C}},\label{GroundStateSiV}\\
|\Psi_{ux,uy}^{(0)} \rangle &=& {1 \over \sqrt{1 + 2\mathcal{N} \beta +  \beta^2}} \left(e_{ux,uy}^{\mbox{\scriptsize  C}} + \beta p_{x,y}^{\mbox{\scriptsize  Si}}\right),  \label{ExcitedStateSiV}
\end{eqnarray}
where $e_{gx,gy}^{\mbox{\scriptsize  C}}$ (gerade) and $e_{ux,uy}^{\mbox{\scriptsize  C}}$ (ungerade) are $sp^3$ linear combinations of single electron orbitals associated to the carbon atoms \cite{20}, $p_{x,y}^{\mbox{\scriptsize  Si}}$ are $p_{x,y}$ orbitals associated to the silicon atom (see Figure~\ref{fig:ConfigurationalDiagram}), $\beta$ is a coefficient that indicates the contribution of the latter orbitals and it is estimated  to be $\approx 0.13$ by ab initio calculations, and $\mathcal{N} = \langle p_{x,y}^{\mbox{\scriptsize  Si}} | e_{ux,uy}^{\mbox{\scriptsize  C}} \rangle$. Thanks to inversion symmetry the excited and ground state can also be labeled by parity. The degenerate ground states $|\Psi_{gx}^{(0)} \rangle$ and $|\Psi_{gy}^{(0)} \rangle$ belong to the two-fold IR $E_{g} = \{E_{gx}, E_{gy}\}$, respectively. Meanwhile, the degenerate excited states $|\Psi_{ux}^{(0)} \rangle$ and $|\Psi_{uy}^{(0)} \rangle$ belong to the two-fold IR $E_{u} = \{E_{ux}, E_{uy}\}$, respectively. These ground and excited states are energetically separated by the zero-phonon line energy $E_{\mbox{\scriptsize  ZPL}} = 1.68$ eV \cite{33}. Therefore, the electronic structure associated to the negatively charged SiV$^{-}$ is modeled by the following Hamiltonian
\begin{equation}
H_{\mbox{\scriptsize  e}} = {1 \over 2}E_{\mbox{\scriptsize  ZPL}} \left(|\Psi_{ux}^{(0)}\rangle\langle \Psi_{ux}^{(0)}| - |\Psi_{gx}^{(0)}\rangle\langle \Psi_{gx}^{(0)}| \right).
\label{ESiV}
\end{equation}
We do not include the effect of spin-orbit interaction, neither we include the spin degree of freedom as they are not relevant  for determining the broad features 
of the optical lineshape.

\section{Electron-phonon Hamiltonian}
In this section we derive a model for the electron-phonon interaction between a single SiV$^{-}$ center and lattice vibrations in a finite sized crystalline structure. 
First, we consider a diamond lattice composed of $N_{\mbox{\scriptsize  Lat}}$ atoms including the SiV$^{-}$ center at the origin. 
Atoms are arranged so that the whole structure maintains the $C_{3v+i}$ symmetry of the point defect. 
We introduce the normal coordinates that describe lattice vibrations \cite{1}
\begin{equation}
Q_l^{\mbox{\scriptsize  Lat}} = \sum_{i = 1}^{N_{\mbox{\scriptsize  Lat}}}\sum_{\alpha = \{x,y,z\}}^{}\sqrt{M_i}u_{i \alpha} h_{i \alpha, l}^{\mbox{\scriptsize  Lat}},       \label{NormalCoordinatesLattice}                                                         
\end{equation}
where $M_{i}$ is the mass of the $i$-th ion and $u_{i \alpha}$ is the displacement of the $i$-th ion in the $\alpha$ direction ($x,y$ or $z$). 
In this notation, $\mathbf{u}_i$ is the ion displacement vector from its equilibrium position $\mathbf{R}_i^{(0)}$, and 
$h_{i\alpha, l}^{\mbox{\scriptsize  Lat}}$ are eigenvectors that satisfy the following eigenvalue equation \cite{1}
\begin{equation}
 \sum_{j=}^{N_{\mbox{\scriptsize  Lat}}}\sum_{\beta = \{x,y,z\}}D_{i\alpha, j\beta}h_{j\beta, l}^{\mbox{\scriptsize  Lat}} = \omega_l^2 h_{i\alpha,l}^{\mbox{\scriptsize  Lat}},   
 \; \hspace{0.5 cm} \; l = 1,...,3N_{\mbox{\scriptsize  Lat}},
 \label{EigenvalueProblem}
\end{equation}
where $D_{i\alpha,j\beta}$ is the dynamical matrix associated with the ion-ion potential interaction and $\omega_l$ are the frequency associated with the $l$-th lattice mode. The dynamical matrix is given by \cite{1}
\begin{equation}
 D_{i\alpha ,j\beta} = {1 \over \sqrt{M_i M_j}}\left. \left({\partial^2 V_{\mbox{\scriptsize  Ion-Ion}} \over \partial u_{i\alpha} \partial u_{j\beta}}\right)\right|_{\mathbf{R}_0},
 \label{DynamicaMatrixDij}
\end{equation}
where $V_{\mbox{\scriptsize  Ion-Ion}}$ is the ion-ion Coulomb interaction (see Appendix B for further details). The electron-phonon interaction between the electronic states associated to this point defect and lattice vibrations can be written as
 \begin{equation}
 V_{\mbox{\scriptsize  e-ph}}(\mathbf{r},\{\mathbf{Q}\})  = \sum_{l=1}^{3N_{\mbox{\scriptsize  Lat}}-6} \left[\sum_{l'=1}^{3N_{\mbox{\scriptsize  D}}-6} \alpha_{l'l} \left( {\partial V_{\mbox{\scriptsize  e-Ion}} \over \partial Q_{l'}^{\mbox{\scriptsize  SiV}}} \right) \right] Q_l^{\mbox{\scriptsize  Lat}}, \label{EphInteraction}
\end{equation}
where $N_{\mbox{\scriptsize  D}}$ is the number of defect atoms ($N_{\mbox{\scriptsize  D}} = 7$ for the SiV$^-$ center), $V_{\mbox{\scriptsize e-Ion}}$ is the electron-ion Coulomb interaction between one electron located at $\mathbf{r}$ and the $N_{\mbox{\scriptsize Lat}}$ surrounding atoms, and $Q_{l'}^{\mbox{\scriptsize SiV}}$ are the local 
normal coordinates of the SiV$^{-}$ center. The factor $\alpha_{l'l}$ is given by
\begin{equation}
 \alpha_{l'l} =  \langle \mathbf{H}_{l'}^{\mbox{\scriptsize SiV}}, \mathbf{h}_{l}^{\mbox{\scriptsize Lat}} \rangle = 
 \sum_{i=1}^{N_{\mbox{\scriptsize  D}}}\sum_{\alpha=\{x,y,z\}} H_{i \alpha, l'}^{\mbox{\scriptsize SiV}} \; h_{i\alpha, l}^{\mbox{\scriptsize Lat}},
 \label{ProjectionLocalOverLatMode}
\end{equation}
where $\mathbf{H}_{l'}^{\mbox{\scriptsize SiV}}$ center and $\mathbf{h}_{l}^{\mbox{\scriptsize Lat}}$ are the eigenvectors associated to the vibrational modes of the 
SiV$^-$ and the finite lattice structure. We assume that the electron wavefunctions are non-zero only on the $N_{\mbox{\scriptsize  D}}$ defect atoms, therefore it is sufficient to consider the inner sum on the defect atoms only. In the Appendix A we show a full derivation of the electron-phonon interaction. 
Next, we promote the normal coordinates and the corresponding momentum conjugate to operators as follows 
\begin{equation}
Q_{l}^{\mbox{\scriptsize Lat}} = \sqrt{\hbar \over 2 \omega_l}\left(\hat{b}_l^{\dagger} + \hat{b}_l  \right),
\; \hspace{0.5 cm} \; P_{l}^{\mbox{\scriptsize Lat}} = i \sqrt{\hbar \over 2 \omega_l}\left(\hat{b}_l^{\dagger} - \hat{b}_l \right),
\label{QuantizationQ}
\end{equation}
where the set of $3N_{\mbox{\scriptsize Lat}}-6$ independent boson creation $\hat{b}_l^{\dagger}$ and 
annihilation $\hat{b}_l$ operators obey the commutation relation
\begin{equation}
[\hat{b}_l,\hat{b}_{l'}^{\dagger}] = \delta_{ll'}.
\end{equation}
Note that we only quantize vibrational modes, as translational and rotational modes leave invariant 
the electron-phonon interaction. Finally, by expanding the electron-phonon interaction in the electronic basis $| i \rangle = \{|\Psi_{gx}^{(0)} \rangle,|\Psi_{ux}^{(0)} \rangle \}$ the following electron-phonon Hamiltonian is obtained
\begin{equation}
H_{\mbox{\scriptsize  e-ph}} = \sum_{i,l}^{} \lambda_{i,l} |i \rangle \langle i| (\hat{b}_l^{\dagger} + \hat{b}_l), 
\label{HamiltonianElectronPhononInteractionFirstOrder}
\end{equation}
where the electron-phonon coupling constants are given by
\begin{eqnarray}
\lambda_{i,l} &=& \sqrt{\hbar \over 2\omega_l}\sum_{l'=1}^{3N_{\mbox{\scriptsize D}}-6}\langle \mathbf{H}_{l'}^{\mbox{\scriptsize SiV}}, \mathbf{h}_{l}^{\mbox{\scriptsize Lat}} \rangle  \gamma_{i,l'} \label{Lambda}
\\
\gamma_{i,l'} &=& \langle i |\left. \left(  {\partial V_{\mbox{\scriptsize e-Ion}} \over \partial Q_{l'}^{\mbox{\scriptsize SiV}}} \right) \right|_{\mathbf{R}_0} | i \rangle.
\label{Electron-Phonon-Coupling}
\end{eqnarray}
To evaluate $\gamma_{i,l'}$ we used symmetrized Gaussian orbitals (see Appendix C for details). 
On Eq.\eqref{HamiltonianElectronPhononInteractionFirstOrder} we have only kept those terms that shift the energy of the electronic states. Other terms such as
\begin{equation} 
\sum_{i \neq j,l}\lambda_{ij,l} | i \rangle \langle j | (\hat{b}_l^{\dagger} + \hat{b}_l),
\end{equation}
are not considered. The latter terms make Hamiltonian \eqref{HamiltonianElectronPhononInteractionFirstOrder} analytically unsolvable for a direct diagonalization calculation \cite{6}. Nevertheless, these terms will be considered by means of dynamical symmetry breaking.

\section{Model for the emission spectrum}
The fluorescence spectrum of the emitted radiation in a thermal equilibrium state is determined by the spectral intensity radiated per unit 
solid angle by an oscillating dipole and it is given by \cite{11}
\begin{eqnarray}
{dI \over d\Omega} &=& {\omega_0^4 \over 8 \pi^2 c^3} \left|\left(\mathbf{n} \times \mathbf{d} \right) \times \mathbf{n} \right|^2 \; E(\omega),
\\
E(\omega) &=& \int_{-\infty}^{\infty} \langle \sigma_{-}(t) \sigma_{+}(0) \rangle_{\mbox{\tiny eq}} \; e^{-i\omega t} \, dt ,
\end{eqnarray}
where $\mathbf{d}$ is the dipole vector and $\mathbf{n} = \mathbf{r}/|\mathbf{r}|$ is the unitary vector pointing 
in the direction of $\mathbf{r}$. Therefore, we calculate the emission spectrum associated to the electronic transition from the excited $|e\rangle$ to ground state $|g\rangle$ as the Fourier transform of the current-current correlation function at thermal equilibrium by applying the Kubo formula \cite{7,11}
\begin{equation}
E(\omega) = \int_{-\infty}^{\infty} e^{-i\omega t} \langle \sigma_{-}(t) \sigma_+(0)\rangle_{\mbox{\scriptsize  eq}}\; dt, \label{EmissionSpectrum1}
\end{equation}
where $\sigma_{+} = | e\rangle\langle g|$, $\sigma_{-} = | g \rangle \langle e |$, $\sigma_{\pm}(t)= U^{\dagger}(t)\sigma_{\pm}(0)U(t)$ and $U(t) = e^{-i H_{\mbox{\scriptsize  SB}}t/\hbar}$. The Hamiltonian $H_{\mbox{\scriptsize  SB}}$, known as the spin-boson Hamiltonian \cite{7}, is given by
\begin{equation}
H_{\mbox{\scriptsize  SB}} = H_{\mbox{\scriptsize  e}} + H_{\mbox{\scriptsize  e-ph}}  + \sum_{l} \hbar \omega_l \hat{b}_l^{\dagger}\hat{b}_l, 
 \label{Hamiltonian}
\end{equation}
where the first, second and third term are the Hamiltonians of the electronic states of the point defect (Eq.\eqref{ESiV}), the electron-phonon interaction to first order in the 
ion displacements (Eq.\eqref{HamiltonianElectronPhononInteractionFirstOrder}), and the phonon bath, respectively. The average 
$\langle ... \rangle_{\mbox{\scriptsize eq}}$ is taken over phonons, which are assumed to be in thermal equilibrium. 
The electron-phonon interaction in Eq.\eqref{Hamiltonian} describes acoustic, optical and quasi-local phonon modes coupled to the electronic states of the point defect. Physically, during the emission or absorption processes, the electronic charge changes its spatial distribution leading to a change in the potential seen by the ions close to the charge localization. Ions will seek for new equilibrium positions, resulting in a relaxation process inducing lattice vibrations. In order to determine 
how the phonon relaxation processes affect the optical properties we introduce the polaron transformation \cite{7,12} given by 
\begin{equation}
H' = e^{S}  He^{-S},
\end{equation}
where 
\begin{equation}
S = \sum_{i,l} {\lambda_{i,l} \over \hbar \omega_l} | i \rangle \langle i | \left(\hat{b}_l^{\dagger} - \hat{b}_l \right).
\end{equation}
In the density operator formalism, the state of thermal equilibrium 
that maximizes the von Neumann entropy $S(\hat{\rho}) = -\mbox{Tr}\left(\hat{\rho}\ln \hat{\rho} \right)$ is given by 
$\hat{\rho}_{\mbox{\tiny eq}} = e^{-\beta H_{\mbox{\tiny  SB}}} / Z$, where $Z = \mbox{Tr}\left(e^{-\beta H_{\mbox{\tiny  SB}}} \right)$ is the 
partition function, $\beta = 1/k_{\mbox{\tiny B}}T$, $T$ is the temperature, and $k_B$ is the Boltzmann constant. Therefore, the expectation value 
can be calculated as
\begin{equation}
\langle \sigma_{-}(t) \sigma_+(0)\rangle_{\mbox{\scriptsize  eq}}  = 
{1 \over Z}  \mbox{Tr}\left(\sigma_{-}'(t) \sigma_+'(0) e^{-\beta H_{\mbox{\tiny  SB}}' } \right).
\end{equation}
Under these approximations the emission spectrum can be analytically calculated as
\begin{eqnarray}
E(\omega) = \int_{-\infty}^{\infty} \; e^{-i(\omega-\omega_{eg}+\Delta_e-\Delta_g) t + \Phi(t)} \, dt,
\end{eqnarray}
where $\omega_{eg} = \omega_e - \omega_g$ is the bare electronic frequency transition, $\Delta_{i} = \sum_{l} \lambda_{i,l}^2/\hbar\omega_l$ is the polaron shift and 
$\Phi(t)$ contains the effect of phonons on the optical line shape and is given by
\begin{flalign}
\Phi(t) &= \int_{0}^{\infty} {J_0(\omega) \over \left(\hbar\omega \right)^2}\left[\coth\left({\beta \hbar \omega \over 2} \right) \left(\cos\omega t - 1 \right) - i \sin \omega t \right] \; d \omega,&
\end{flalign}
and 
\begin{equation}
J_0(\omega) =  \sum_{l}\left(\lambda_{e,l}-\lambda_{g,l} \right)^2 \delta(\omega-\omega_l), \label{SpectralFunction}
\end{equation}
is the spectral density function where $\lambda_{i,l}$ is the expectation value of the electron-phonon coupling between phonon modes $l$ and the electronic wavefunction $|i \rangle$. If the electronic states interact with the same strength to phonons, both coupling constants for the ground and excited states are similar and the spectral density function is small leading to a transition involving few phonons and resulting in a fluorescent shape that closely resembles that of a phonon-free system. On the contrary, if these two couplings are substantially different, the change on electronic distribution, and therefore, on the potential seen by the ions is large and the emission spectrum is greatly modified (Figure~\ref{fig:ConfigurationalDiagram}).

\begin{center}
\includegraphics[width= 0.8 \linewidth]{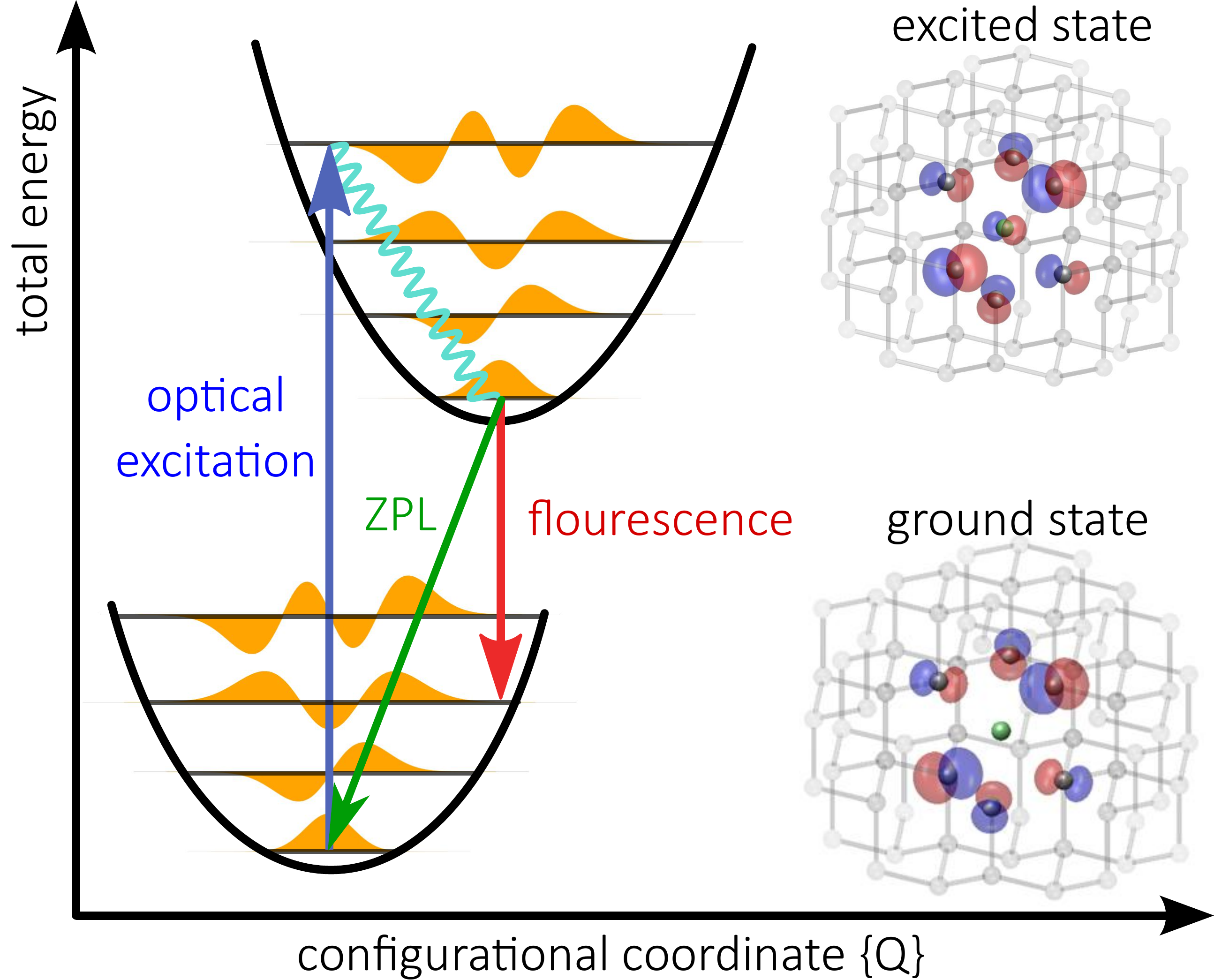}
\captionof{figure}[]{\textbf{Schematic representation of the potential energy diagram}. The two parabolas represent the phononic potential of the ground $e_{gx}$ and excited $e_{ux}$ states of the SiV$^-$  including vibrational levels. Structure of the SiV$^-$ in diamond:  six carbon atoms (dark gray) and the interstitial silicon atom (green) embedded in a diamond lattice (light gray) . The molecular orbital representation of the electronic states $e_{gx}$ and $e_{ux}$ are represented by red (blue) for the positive (negative) sign of the electronic wavefunction.}
\label{fig:ConfigurationalDiagram}
\end{center}

\section{Role of inversion symmetry on the emission spectrum}
The electron-phonon coupling constants depend crucially on the atomic configuration, the symmetry of the point defect and the symmetry of the host material. As an example, the fluorescent of the nitrogen-vacancy center (NV-center) and SiV$^-$ center in diamond are very different from each other although they differ in one atom in their molecular composition. The NV-center has a broad emission ranging from 637 nm zero-phonon line (ZPL) to 750 nm, meanwhile the emission of the SiV$^-$ has a width of few nanometers at the same temperature \cite{13}. The symmetry of the point defect is determined by the atomic configuration \cite{14}. In the case of the NV-center, the nitrogen atom is substitutional and its atomic configuration does not remain the same under inversion, i.e., parity is not a good description for wavefunctions and vibrations \cite{15}. On the contrary, in the SiV$^-$, the silicon atom is interstitial between two vacancies and its configuration remains the same under inversion \cite{16}, i.e., electronic wavefunctions and vibrations can be described by parity. As the coupling constants $\lambda_{i,l}$ are the integration of three functions, its expectation value will be zero if the total product is odd. The lack of inversion symmetry in the NV-center allows in principle the contribution from all vibrational modes. Whereas the coupling constants $\lambda_{e,l}$ and $\lambda_{g,l}$ for the SiV$^-$ can be very similar due to inversion symmetry. Indeed, in the SiV$^-$ the ground state is a gerade (even) linear combination of dangling bond atomic orbitals meanwhile the excited state is an ungerade (odd) function of these orbitals. These wavefunctions might differ only by a phase leading to a very similar electronic 
distribution, a small change upon electronic transitions in the trapping potential seen by the ions, and therefore a very small phonon contribution to the spectral density function $J_0(\omega)$. 

\section{Spectral density function and the emission spectrum}
A quantitative analysis of the phonon modes can be performed by considering a macro molecule composed of $N \sim 10^3$ atoms where the defect is placed at its center 
as described in Section III. The vibrational modes are calculated using a force-constant model to second order nearest-neighbor interaction \cite{17,18} in order to better resemble the real phonon dispersion relation of diamond \cite{19,29} (see Figure~\ref{fig:PhononRelationDispersionDiamond}). Using only a first order nearest-neighbor model does not give an accurate description of the high density areas for the acoustic bands from which arouses the main contribution to the spectral density function. In the Appendix B we show the numerical methodology implemented to obtain the vibrational properties of the macromolecule. 

\begin{center}
\includegraphics[width= 0.9 \linewidth]{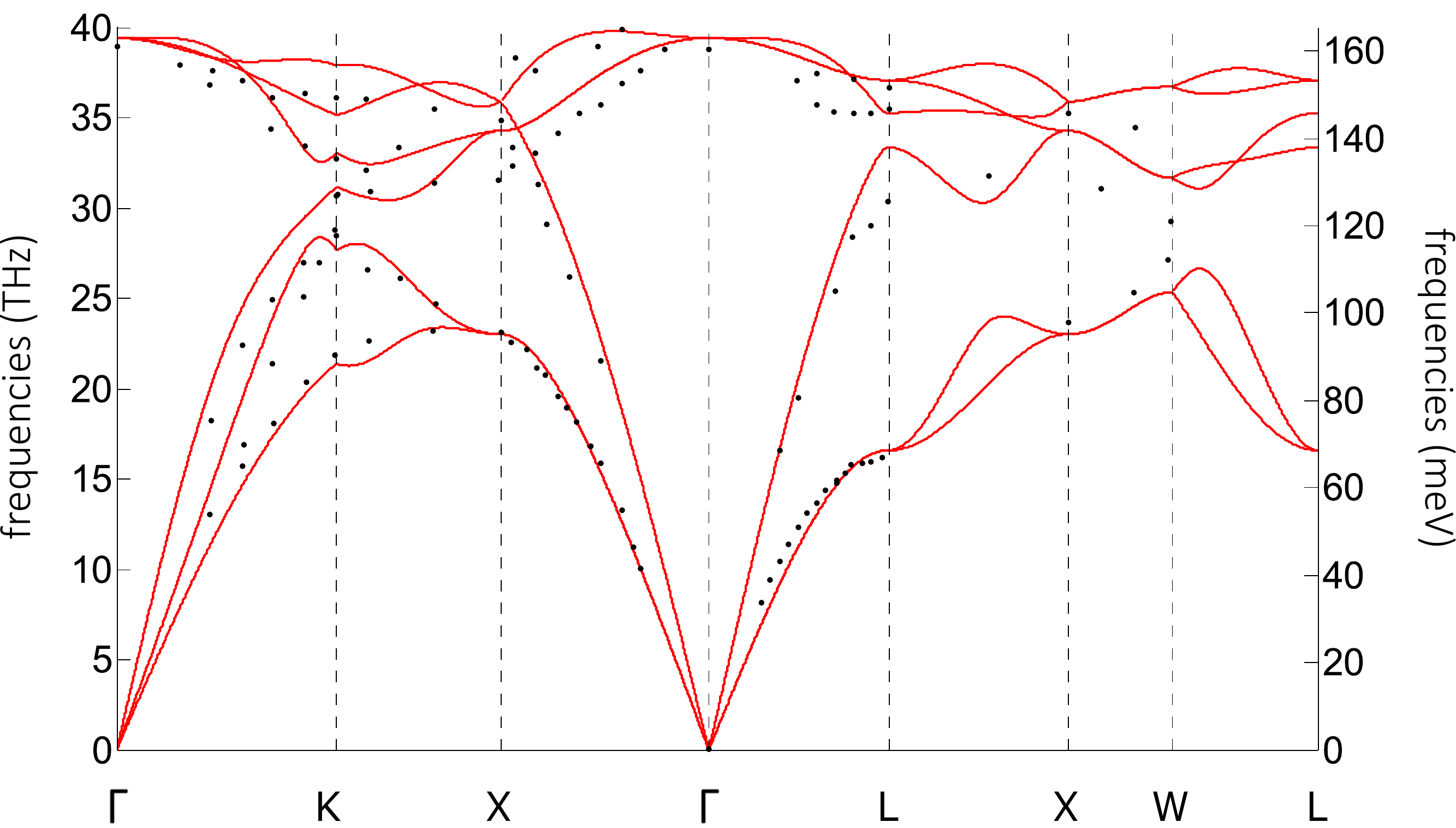}
\captionof{figure}[]{\textbf{Numerical phonon dispersion curves for diamond}. Red lines and black circles correspond to the numerical calculations using the force-constant model to second order nearest-neighbor interactions and experimental neutron-scattering data extracted from \cite{19}. The phonon frequencies are plotted as a function of the reduced phonon wave-vector between some symmetry points in the First Brillouin Zone.}
\label{fig:PhononRelationDispersionDiamond}
\end{center}

Vibrational modes of even parity ($a_{1g}, a_{2g}$ and $e_g$ phonons) contribute to the spectral density function $J_0(\omega)$ associated to the transition $| \Psi_{ux}^{(0)} \rangle \longrightarrow | \Psi_{gx}^{(0)} \rangle$ (see Figure~\ref{fig:SpectralFunctions}a) with the breathing mode of symmetry $a_{1g}$ being the strongest contribution. This peak also contains contributions from $e_g$ phonon modes which contribute to the width of the peak. So far the motion of the silicon atom does not play a role if we consider phonon modes with even symmetry. However, recently an isotopic shift of the phonon sideband was observed for different silicon isotopes \cite{21}: as the mass of the silicon atom increases, the distance between the ZPL and the phonon sideband decreases suggesting that a local vibrational mode primarily composed of the silicon atom is involved. Such mode is necessarily of character $u$ (odd), and for symmetry reasons it should not contribute to the coupling constants $\lambda_{e,l}$ and $\lambda_{g,l}$ if the electronic states given in Eqs.\eqref{GroundStateSiV}-\eqref{ExcitedStateSiV} are used. This indicates that inversion symmetry is broken and it is no longer a good description of the wavefunctions. Inversion symmetry can be broken by vibrations of character $u$, which can dynamically mix both ground and excited states. External electric fields can also break inversion symmetry. Global strain does not mix  ground and excited states as it only mix the states among each manifold \cite{20}. In addition, ab initio calculations support that inversion symmetry is not broken if vibrations are not included. In this scenario, the new electronic wavefunctions can be described by
\begin{eqnarray}
|\Psi_g \rangle &=& \sqrt{1-\epsilon^2} |\Psi_{g}^{(0)} \rangle - \epsilon e^{+i\theta} |\Psi_{e}^{(0)} \rangle \label{WaveGPert} \\
|\Psi_e \rangle &=& \sqrt{1-\epsilon^2} |\Psi_{e}^{(0)} \rangle + \epsilon e^{-i\theta} |\Psi_{g}^{(0)} \rangle \label{WaveEPert},
\end{eqnarray}
 where $\epsilon$ is a mixing parameter, $\theta$ is an arbitrary phase, and $|\Psi_g^{(0)} \rangle$, $|\Psi_{e}^{(0)} \rangle$ are the electronic wavefunctions given in Eqs.\eqref{GroundStateSiV}-\eqref{ExcitedStateSiV}. A similar argument can be given by means of the Herzberg-Teller effect which can also show a dynamical symmetry breaking \cite{22, 23, 24}. The spectral density function $J(\omega) = \sum_{l}\left(\lambda_{\Psi_{e},l} - \lambda_{\Psi_{g},l}\right)^2 \delta(\omega - \omega_l)$ can be explicitly calculated in order to incorporate the effect of the dynamical symmetry breaking given by the mixing of the ground and state states of the SiV$^-$ center. 
 Using group theoretical arguments, averaging over the phase $\theta$ and evaluating in the small mixing limit ($|\epsilon| \ll 1$) we find that \cite{29}
\begin{equation}
J(\omega) = J_0(\omega) + 8 \epsilon^2 J_{eg}(\omega) \label{Totalspectral density function}
\end{equation}
where $J_0(\omega)$ is given by Eq. \eqref{SpectralFunction} and
\begin{equation}
J_{eg} = \sum_{ l} \left( \lambda_{eg,l} \right)^2 \delta(\omega-\omega_l), \; \hspace{0.3 cm} \; \lambda_{eg,l} = \langle \Psi_{g}^{(0)}  | \mathds{H}_{\mbox{\scriptsize  e-ph}}^{(l)} | \Psi_{e}^{(0)} \rangle,
\end{equation}
is the spectral density function that incorporates the contribution of phonon modes with odd symmetry. See Appendix D for a derivation of the spectral density function 
$J_{eg} (\omega)$. Figure~\ref{fig:SpectralFunctions}b shows $J_{eg}(\omega)$ where a strong peak associated to an $a_{1u}$ quasi-local phonon mode is observed with a frequency of $\omega_{28}  = 63.19$ meV, $\omega_{29}  = 62.66$ meV and $\omega_{30}  = 62.16$ meV for isotopes $\tensor[^{28}]{\mbox{Si}}{}$, $\tensor[^{29}]{\mbox{Si}}{}$ and $\tensor[^{30}]{\mbox{Si}}{}$, respectively. The ratio between these energies is approximately $\omega_{28}/\omega_{29} \approx 1.01$ and $\omega_{28}/\omega_{30} \approx 1.02$ and has a good agreement with experimental values ($\omega_{28}/\omega_{29} = 1.016$ and $\omega_{28}/\omega_{30} = 1.036$ \cite{21}). However, the exact value for the energy of this $a_{1u}$ quasi-local phonon mode can be better estimated with more precise methods. The prominent sharp feature of $J_{eg}(\omega)$ has also contributions from $e_u$ and $a_{2u}$ modes where $e_u$ modes contribute approximately twice as much as the $a_{2u}$ modes. The frequency of the 
quasi-local phonon mode $a_{1u}$ has a strong dependence on the silicon mass. In this mode, the silicon atom moves along the symmetry axis. In addition, we observe that $J_{eg}(\omega)$ is considerably larger that $J_0(\omega)$ and strongly depends on the silicon contribution to the electronic wavefunction (see Eq.\eqref{ExcitedStateSiV}). Only a small mixing parameter is sufficient to make $J_{eg}(\omega)$ the largest contribution to the spectral density function given in Eq.\eqref{Totalspectral density function} \cite{29}. 

\vspace{0.1 cm}

This microscopic procedure allows to numerically calculate the contribution of acoustic, optical and quasi-local phonon modes to the spectral density function. However, a large number of atoms is required to have a better estimate of the mode density and of the emission spectrum. Alternatively, known models of the spectral density function can be fitted to simplify the effect of phonons. Bulk phonons have been modelled with a spectral density function of the form \cite{5} $J_{\mbox{\scriptsize  Bulk}}(\omega)  = 2\alpha \omega_c^{1-s} \omega^s e^{-\omega/\omega_c} $, where $\alpha$ is the dissipation strength, $\omega_c$ is a cut-off frequency and $s$ is a dimensionless parameter characterizing the regimes: sub-ohmic ($s<1$), ohmic ($s=1$) and super-ohmic ($s>1$). At low frequencies the contribution from acoustic phonon modes to the SiV$^-$ can be modeled as $J(\omega) \propto \omega^3$ which implies a super-ohmic regime ($s=3$) \cite{9}. For quasi-local phonons $J_{\mbox{\scriptsize  Loc}}(\omega)  = {J_0 \over \pi}{{1 \over 2}\Gamma \over \left(\omega - \omega_b \right)^2 +  \left({1 \over 2}\Gamma \right)^2}$ \cite{20}, where $J_0$ is the coupling strength, $\Gamma$ is a characteristic width and $\omega_b$ is the frequency of the phonon. In the numerical estimation at least two localized contributions $J_{\mbox{\scriptsize  Loc1}}(\omega)$ and $J_{\mbox{\scriptsize  Loc2}}(\omega)$ are recognised at 63.19 meV and around 45.5 meV, respectively. We fit $J_{eg}(\omega)$ to a spectral density function of the form $J_{eg}(\omega) = J_{\mbox{\scriptsize  Bulk}}(\omega)+ J_{\mbox{\scriptsize  Loc1}}(\omega)+ J_{\mbox{\scriptsize  Loc2}}(\omega) $ \cite{25}. We found, however, that $J_{\mbox{\scriptsize  Loc 2}}(\omega)$ is best fit to a gaussian function as it is probably composed of multiple quasi-local phonon modes.
\begin{center}
\includegraphics[width= 0.85 \linewidth]{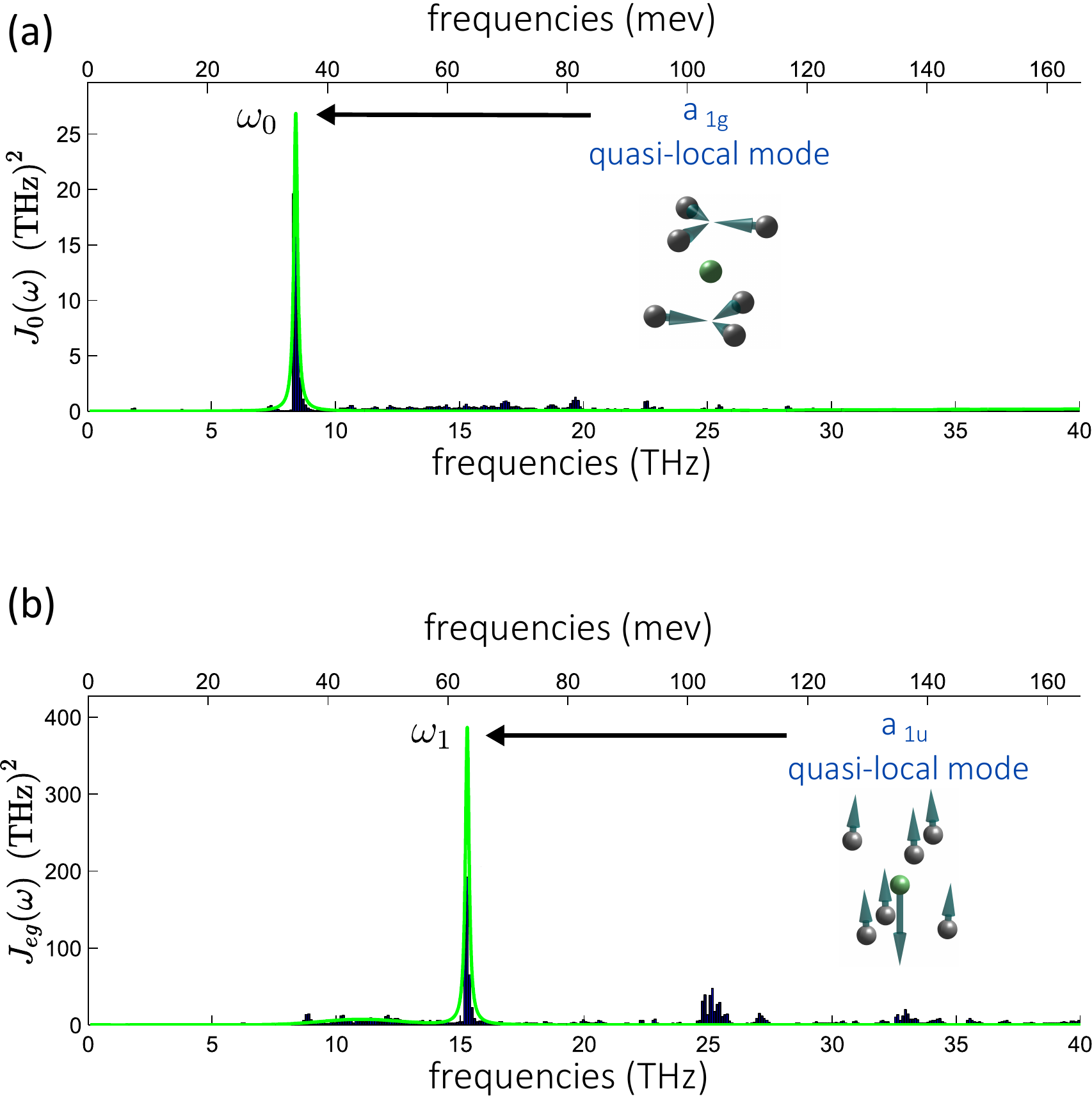}
\captionof{figure}[]{\textbf{Numerical spectral functions $J_0(\omega)$ and $J_{eg}(\omega)$ for the SiV$^-$ in diamond}. (a) Spectral function $J_0(\omega)$, where the blue bar graph and the green line are the numerical estimation and the fit spectral function obtained from simulations. The strongest contribution is given by an $a_{1g}$ phonon mode (breathing mode) at around $\omega_0 = 37$ meV. (b)  Spectral function $J_{eg}(\omega)$, where the blue bar graph and the green line are the numerical estimation and the fit spectral function, respectively. The strongest contribution is given by an $a_{1u}$ quasi-local phonon mode at around $\omega_1=63.19$ meV. A second contribution of the $J_{eg}(\omega)$ spectral function is given at around $\omega_2 = 45.5$ meV.}
\label{fig:SpectralFunctions}
\end{center}
The emission spectrum associated with $J_{eg}(\omega)$ is shown on Figure~\ref{fig:EmissionSpectrum} and has good agreement with the observed isotopic shift \cite{21, 26,27}. The largest contribution to the phonon sideband at 766 nm is due to the main peak in $J_{eg}(\omega)$ at 63.19 meV and it is associated to an $a_{1u}$ quasi-local mode as previously discussed (see Figure 3b). Changing the isotopic mass indeed shifts the distance between the ZPL and this feature on the phonon sideband confirming previous observations \cite{21}. A second contribution to the sideband is observed at 755 nm and is associated with a peak in $J_{eg}(\omega)$ at 45.5 meV and does not have a dependence on the silicon mass. Other peaks in the observed experimental phonon sideband \cite{27} can be associated to other features in the spectral density function $J_0(\omega)$ and $J_{eg}(\omega)$. A peak at 796 nm (with no dependence on the silicon mass) \cite{21} might correspond to the highest phonon frequency of the acoustic band of highest sound speed, close to the L symmetry point of the measured dispersion relation \cite{18,28}.

\vspace{0.1 cm}

Our second nearest-neighbor model over estimate mode frequencies at higher frequencies and locates this points at 136.5 meV, frequency at which there seems to be a contribution on the spectral function $J_{eg}(\omega)$ (see Figure 3b). A similar argument applies for a contribution at 87 meV in the observed phonon sideband corresponding to a 103.4 meV feature in $J_{eg}(\omega)$. The model also allows to calculate temperature effects. As an example, we have plot the emission spectrum at 4K and 297 K (see Figure~\ref{fig:EmissionSpectrum}). Finally, we remark that the isotopic shift is not possible to explain with phonons that transform evenly under inversion. Therefore, a dynamical symmetry breaking is needed, which can be caused by non-inversion preserving perturbations such as external electric fields or odd vibrational modes. 

\vspace{0.1 cm}

Further improvements of the current numerical estimations  can be performed by increasing the number of atoms around the defect for which the defect electronic wavefunctions are non-zero.

\begin{center}
\includegraphics[width=0.9 \linewidth]{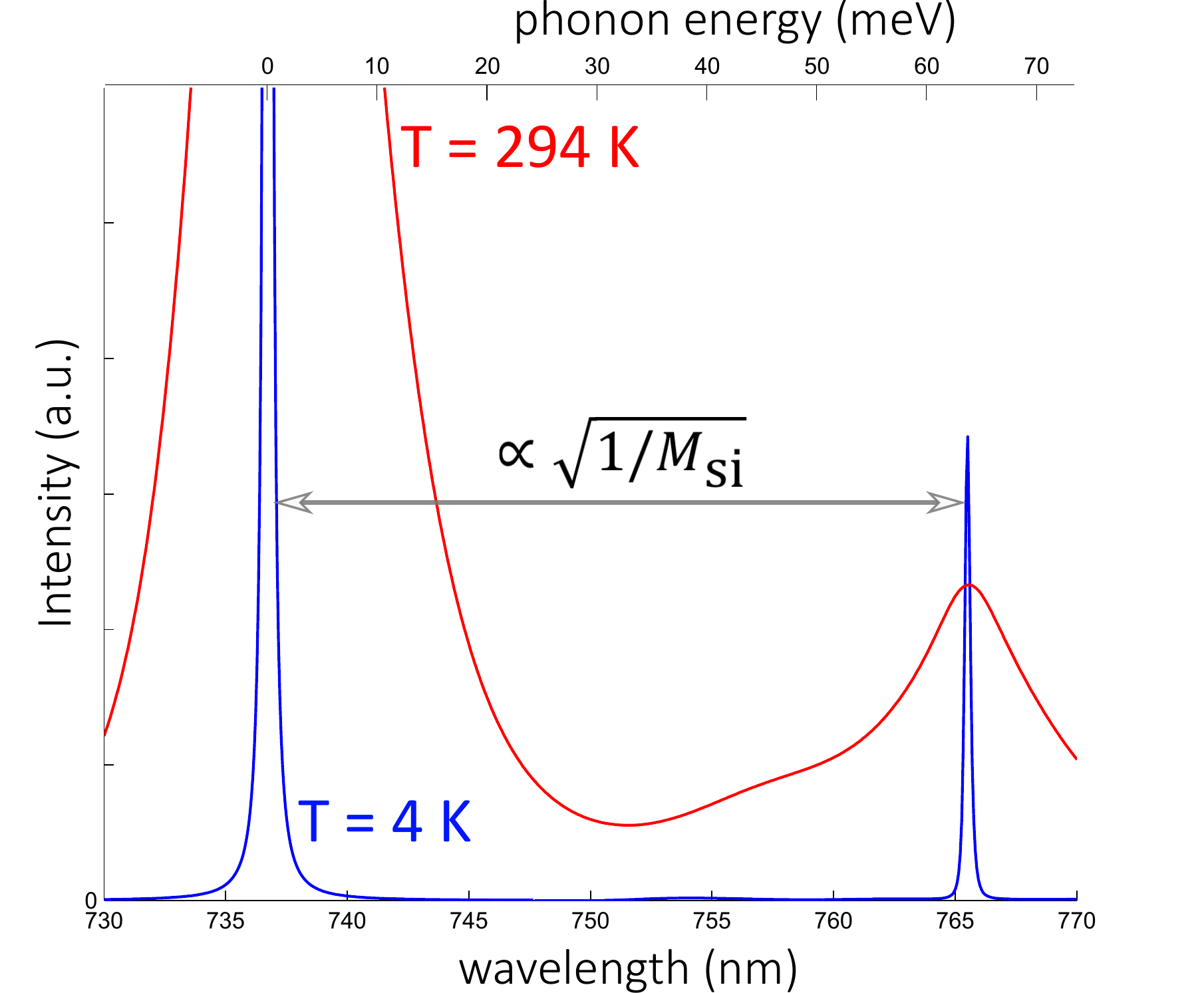}
\captionof{figure}[]{\textbf{Numerical emission spectra of the SiV$^-$ in diamond.}  The blue and red curves represent the numerical emission spectrum obtained  for $T=4$ K and $T=296$ K, respectively. The ZPL at $736$ nm and the prominent sharp feature of the phonon sideband at $766$ nm are reproduced. The peak at $766$ nm its associated with the $a_{1u}$ quasi-local phonon mode. }
\label{fig:EmissionSpectrum}
\end{center}

\section{CONCLUSIONS}
In summary we have presented a microscopic model for estimating the emission spectrum of the SiV$^-$ using the Kubo formula and the spin-boson model. In addition we have considered effects to second-order on the spectral density function via dynamical symmetry breaking. This spectral density function is estimated using a force-constant model for describing the vibrational modes and symmetrized electronic wavefunctions constructed using group theoretical arguments. This approach allows us to gain detailed insight on the microscopic origin and the role of symmetries on the emission spectra and the spectral density function, approach which is crucially different from, but validates, phenomenological models presented in previous works \cite{5,6,8}. These results might be useful for understanding and engineering the optical properties of colour centers in solids by extending the analysis to other deep and shallow centers coupled to phonons and subject to instabilities such as dynamic Jahn-Teller effects and external perturbations such as electric fields or strain.

\section*{ACKNOWLEDGMENTS}
The authors acknowledge fruitful discussions with Marcus Doherty at the Diamond Quantum Sensing 2015. J.R. acknowledge support from Conicyt-Fondecyt 1141185, Conicyt-PIA ACT1108, and AFOSR grant FA9550-15-1-0113. A.N acknowledges support from Conicyt fellowship No. 21130645. J.M acknowledge support from Fondecyt grants No. 1130672 and Basal Funding for scientific and technological centers of excellence BF 0807. AG acknowledges Lendület program of the Hungarian Academy of Sciences and EU FP7 project DIADEMS grant No. 611143.

\section*{Appendix A: Electron-phonon interaction}
In this section we present a more detailed derivation of the electron-phonon interaction used to model the optical properties of the SiV$^-$ center. Using the normal coordinates $Q_l^{\mbox{\scriptsize Lat}}$ defined in 
Eq.\eqref{NormalCoordinatesLattice} the electron-phonon interaction can be expanded as follow
\begin{equation}
 V_{\mbox{\scriptsize e-ph}}(\mathbf{r},\{\mathbf{Q}\}) = V_0 + 
 \sum_{l=1}^{3N_{\mbox{\scriptsize Lat}}-6} \left( {\partial V_{\mbox{\scriptsize e-Ion}} \over \partial Q_l^{\mbox{\scriptsize Lat}}} \right) Q_l^{\mbox{\scriptsize Lat}} + \dots, 
 \label{ExpansionElectronPhononInteraction}
\end{equation}
where only the $3N_{\mbox{\scriptsize Lat}}-6$ vibrational modes are considered, as translational and rotational modes leave invariant 
the electron-phonon interaction \cite{1}. As we will focus on deep centers, \textit{i.e.}, center whose electronic wave functions decay quickly with distance \cite{30}, it will be convenient to define local vibrational modes involving only those atoms on which the electronic wave functions are considered to be non-zero. These modes can be obtained from group theoretical considerations \cite{1,14} or by numerically solving a small molecular system considering only the atoms related with the defect structure using a force-constant model \cite{31} or ab initio calculations. These defect normal coordinates  are defined as
\begin{equation}
Q_{l'}^{\mbox{\scriptsize SiV}} = \sum_{i = 1}^{N_{\mbox{\scriptsize  D}}}\sum_{\alpha = \{x,y,z\}}^{}\sqrt{M_i}u_{i \alpha} h_{i \alpha, l'}^{\mbox{\scriptsize SiV}}, 
\label{LocalNormalCoordinates2}
\end{equation}
where $N_{\mbox{\scriptsize D}}$ is the number of atoms of the defect ($N_{\mbox{\scriptsize D}} < N_{\mbox{\scriptsize Lat}}$), $u_{i \alpha}$ is the displacement of the $i-$th ion in the $\alpha$ direction from its equilibrium position, and $h_{i \alpha, l'}^{\mbox{\scriptsize SiV}}$ are the eigenvectors $l'$ associated to the defect molecular vibrations of the $i-$th ion in the $\alpha$ direction. The local normal coordinates of the defect can be written as a linear combination of the lattice normal modes given in 
Eq. \eqref{NormalCoordinatesLattice}
\begin{equation}
Q_{l'}^{\mbox{\scriptsize SiV}} = \sum_{l=1}^{3N_{\mbox{\scriptsize Lat}}-6}\alpha_{l'l}Q_{l}^{\mbox{\scriptsize Lat}},
\label{LocalNormalCoordinates}
\end{equation}
where the parameter $\alpha_{l' l}$ is given by Eq.\eqref{ProjectionLocalOverLatMode}. $\mathbf{H}_{l'}^{\mbox{\scriptsize SiV}}$ and $\mathbf{h}_{l}^{\mbox{\scriptsize Lat}}$ are vectors with the same dimensionality and whose components are given by
\begin{equation}
\mathbf{H}_{l'}^{\mbox{\scriptsize SiV}} = \left( \begin{array}{c}  
                                     h_{1 x, l'}^{\mbox{\scriptsize SiV}}  \\
                                      h_{1 y, l'}^{\mbox{\scriptsize SiV}}  \\
                                       h_{1 z, l'}^{\mbox{\scriptsize SiV}}  \\
                                     \vdots \\
                                     h_{N_{\mbox{\tiny D}} z, l'}^{\mbox{\scriptsize SiV}}  \\
                                     0 \\
                                    \vdots \\ 
                                    0
                                   \end{array} \right), \hspace{0.5 cm} \mathbf{h}_{l}^{\mbox{\scriptsize Lat}} = \left( \begin{array}{c}  
                                     h_{1 x, l}^{\mbox{\scriptsize Lat}}  \\
                                      h_{1 y, l}^{\mbox{\scriptsize Lat}}  \\
                                       h_{1 z, l}^{\mbox{\scriptsize Lat}}  \\
                                     \vdots \\
                                     h_{N_{\mbox{\tiny D}} z, l}^{\mbox{\scriptsize Lat}}  \\
                                     h_{N_{\mbox{\tiny D}}+1 z, l}^{\mbox{\scriptsize Lat}} \\
                                    \vdots \\ 
                                    h_{N_{\mbox{\tiny Lat}} z, l}^{\mbox{\scriptsize Lat}}
                                   \end{array} \right) \label{NewLocalNormalModes}
\end{equation}
where $H_{i\alpha, l'}^{\mbox{\scriptsize SiV}}$ are obtained from group theoretical arguments and $H_{i\alpha, l'}^{\mbox{\scriptsize Lat}}$ are numerically obtained by solving  the eigenvalue equation \eqref{EigenvalueProblem}. Therefore, using the chain rule and neglecting the constant term $V_0$ on Eq.\eqref{ExpansionElectronPhononInteraction} we recover electron-phonon interaction given in Eq.\eqref{EphInteraction}.

\section*{Appendix B: force constant model to second order nearest-neighbor}
\begin{figure}
\begin{center} 
\includegraphics[width= 0.4 \textwidth]{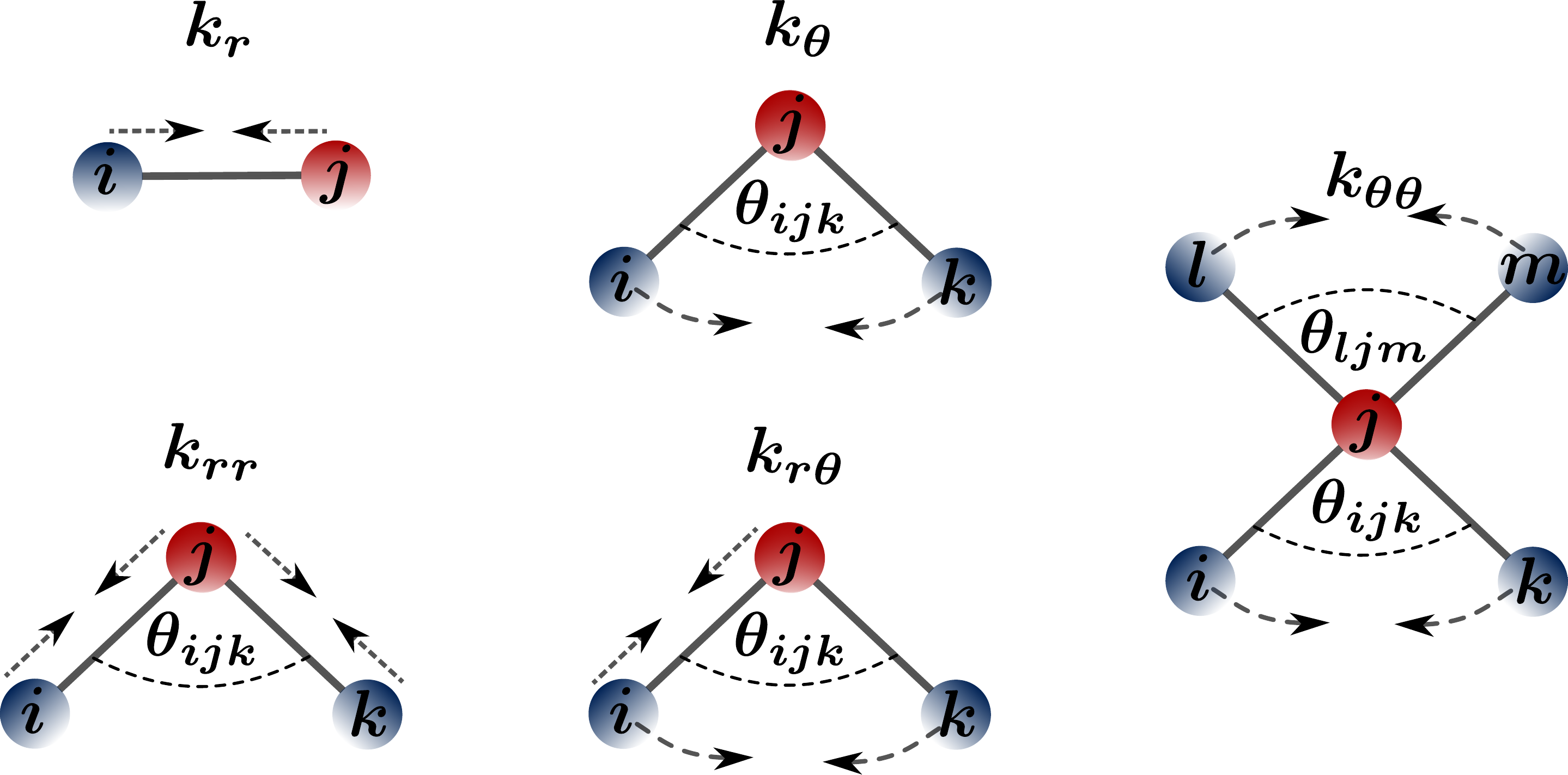}
\caption{\textbf{Types of interactions present in a second nearest neighbor model for phonons}.}
\label{fig:SecondNearestNeighbourInteractions}
\end{center}
\end{figure}

In this section we present the force constant model used to numerically solve the vibrational modes associated to the eigenvalue equation given in \eqref{EigenvalueProblem}. Using the general valence force field for diamond \cite{17}, we can extract the vibrational dynamics of the system using the 
following expression for the ion-ion interaction including up to second nearest-neighbor interactions
\begin{equation}
 V_{\mbox{\tiny Ion-Ion}}  = \sum_{k_s\in \mathds{K}} V_{k_s}, \; \hspace{0.5 cm} \; \mathds{K} =  \{k_r,k_{rr},k_{r\theta},k_{\theta},k_{\theta\theta} \},
 \label{Ion-Ion-Interaction}
\end{equation}
where the contributions to the ion-ion potential interaction are given by
\begin{eqnarray}
V_{k_r} & = &  {1 \over 2} k_r \sum_{\langle ij \rangle} \left( \delta u_{ij} \right)^2, \label{Vkr}  \\
V_{k_{rr}} & = &   k_{rr} \sum_{\langle ij \rangle,\langle kj \rangle} \left( \delta u_{ij} \right)\left( \delta u_{kj} \right),  \label{Vkrr} \\
V_{k_{r\theta}} & = & b k_{r\theta} \sum_{\langle ijk \rangle} \left(\delta u_{ij} \right) \left(\delta \theta_{ijk} \right),  \label{Vkrt}  \\                 
V_{k_{\theta}} & = & {1 \over 2} b^2 k_{\theta} \sum_{\langle ijk \rangle} \left( \delta \theta_{ijk} \right)^2,  \label{Vkt} \\                
V_{k_{\theta \theta}} & = &  {1 \over 2}b^2 k_{\theta \theta}   \sum_{\langle ijk  \rangle , \langle  ljm \rangle} \left( \delta \theta_{ijk} \right)\left( \delta \theta_{ljm} \right) ,  \label{Vktt}                
\end{eqnarray}
These interaction terms are illustrated in Figure \ref{fig:SecondNearestNeighbourInteractions} and depends on the geometrical distortions of the lattice 
\begin{eqnarray}
 \delta u_{ij} &=& |\mathbf{u}_i - \mathbf{u}_j| , \; \hspace{0.5 cm} \; \mathbf{\hat{u}}_{ij} = (\mathbf{u}_i - \mathbf{u}_{j})/\delta u_{ij} \\
 \delta \theta_{ijk} &=&  \cos^{-1}(\mathbf{\hat{u}}_{ij} \cdot \mathbf{\hat{u}}_{kj}),
\end{eqnarray}
and the elastic constants $k_r,k_{rr},k_{r\theta},k_{\theta},k_{\theta\theta}$. These elastic constants are obtained from literature in the case of bulk-diamond 
\cite{17,18} and from ab initio simulations for the SiV$^{-}$ center.  The parameter $b = 1.95$ \AA \; for the point defect and $b = 1.54$ \AA \; for the bulk diamond.
We use the following elastic constants for the SiV$^-$ center
\begin{eqnarray}
k_r^{\mbox{\tiny SiV}}  &=&  45 \;  \mbox{N} / \mbox{m} =  2.8087   \; \mbox{eV}/ \mbox{\r{A}}^{2} \\
k_{rr}^{\mbox{\tiny SiV}}  &=&  17.7 \; \mbox{N} / \mbox{m} = 1.1047  \; \mbox{eV}/\mbox{\r{A}}^{2}\\
k_{r\theta}^{\mbox{\tiny SiV}} &=&  37.5 \; \mbox{N} / \mbox{m} = 2.3406 \; \mbox{eV}/\mbox{\r{A}}^{2}\\
k_{\theta \theta}^{\mbox{\tiny SiV}} &=& 3.5 \; \mbox{N} / \mbox{m} = 0.2091  \; \mbox{eV}/\mbox{\r{A}}^{2}\\
k_{\theta}^{\mbox{\tiny SiV}} &=& 47.23 \; \mbox{N} / \mbox{m} = 2.9479 \; \mbox{eV}/\mbox{\r{A}}^{2}
\end{eqnarray}

\section*{Appendix C: symmetrized Gaussian orbitals and electron-phonon coupling constants}
The electron-phonon coupling constants given in Eqs.\eqref{Lambda}-\eqref{Electron-Phonon-Coupling} can be numerically solved by estimating the following integral
\begin{equation}
\langle i |\left. \left(  {\partial V_{\mbox{\scriptsize e-Ion}} \over \partial u_{i \alpha} } \right) \right|_{\mathbf{R}_0} | j \rangle = 
\int \limits_{\mathds{R}^3} \varphi_{i}^{\ast}(\mathbf{r})   \left. \left(  {\partial V_{\mbox{\scriptsize e-Ion}} \over \partial u_{i \alpha} } \right) \right|_{\mathbf{R}_0}
         \varphi_{j}(\mathbf{r})\; d \mathbf{r}, \label{Integral}
\end{equation}
where the electron-Ion potential is modeled by a screening Coulomb potential  given by
\begin{equation}
V_{\mbox{\tiny e-Ion}} = -\sum_{i = 1}^{N_{\mbox{\scriptsize D}}} {k_{e} Z_i e^2 \over \varepsilon_{\mbox{\tiny D}}|\mathbf{r} - \mathbf{R}_{i}|}, \; \hspace{0.3 cm} \; \mathbf{R}_{i} = \mathbf{R}_{i}^{(0)} + \mathbf{u}_{i},
\end{equation}
where $k_e = 1/(4\pi \varepsilon_0 )$ is the Coulomb constant, $\varepsilon_{\mbox{\tiny D}} = 10$ is the diamond dielectric constant, and the effective charge $Z_i = 3.25, 4.15$ for carbon and silicon atoms, respectively. The electronic wavefunctions $\varphi_{i}(\mathbf{r})$  are approximated by symmetrized Gaussian orbitals in order to numerically solve the integral \eqref{Integral}. In this approximation, the single atomic orbitals for the carbon and silicon atoms are written as linear combinations of the following Gaussian 
orbitals,
\begin{flalign}
s_a &=  \left({2a \over \pi} \right)^{3/4} \exp \left(-a|\mathbf{r} - \mathbf{r}_a|^2  \right), \label{GaussianOrbitalS}  & \\
p_{ak}  &= \sqrt{4 \pi} \left({2a \over \pi} \right)^{3/4} \mathbf{e}_k \cdot  (\mathbf{r} - \mathbf{r}_a) \exp \left(-a|\mathbf{r} - \mathbf{r}_a|^2  \right),& 
\label{GaussianOrbitalP}
\end{flalign}
where $\mathbf{e}_k = \{\mathbf{\hat{x}},\mathbf{\hat{y}},\mathbf{\hat{z}}\}$ for $k=\{x,y,z\}$. The integral \eqref{Integral} can be numerically solved using spherical coordinates 
$(r,\theta,\phi)$ and the seed integral is given by 
\begin{eqnarray}
\int \limits_{\mathds{R}^3} {1 \over r} \exp\left(-a|\mathbf{r}-\mathbf{A}|^2 \right)
                            \exp\left(-b|\mathbf{r}-\mathbf{B}|^2 \right) \; d\mathbf{r}   = S \; { \mbox{erf}(\sqrt{c}\; u) \over u}, \nonumber\\
                            \label{GaussianSeed}
\end{eqnarray}
where 
\begin{eqnarray}
S &=& \left( {2 \sqrt{ab} \over a + b} \right)^{3/2} \exp\left(-{ab \over a+b}|\mathbf{A}-\mathbf{B}|^2 \right), \\
c &=& a+b, \; \hspace{1 cm} \; u = {a|\mathbf{A}| + b|\mathbf{B}| \over a+b}, \\
\mbox{erf}(x) &=& {2 \over \sqrt{\pi}}\int \limits_{0}^{x}e^{-t^2}\; dt.
\end{eqnarray}
Note that integrals involving p-orbitals can be obtained by taking the derivative of equation \eqref{GaussianSeed} with respect to some of the components of the ion positions $\mathbf{A}$ or $\mathbf{B}$. The exponential decay constants of the Gaussian orbitals \eqref{GaussianOrbitalS} and \eqref{GaussianOrbitalP} are determined 
by minimizing the error on the radial probability distribution with respect to the radial probability distribution of the Slater orbitals. We obtain $a = 1.7105$ \AA$^{-2} \;$ for the carbon atoms and $a = 2.9879$ \AA$^{-2} \;$ for the silicon atom.

\section*{Appendix D: Dynamical symmetry breaking and spectral density function} 
In this section we derive the modified spectral density function due to dynamical symmetry breaking. Let $V(t) = V_ue^{-i\omega_{\mbox{\tiny ph}}t}$ be a periodic time-dependent operator which perturbs the localized electronic degree of freedom of SiV$^-$ center. Using time dependent perturbation theory we can define the electronic wavefunctions given in Eqs.\eqref{WaveGPert}-\eqref{WaveEPert}. As a consequence of the mixing effect induced by this external perturbation the effective electron-phonon coupling must be calculated as follows
\begin{equation}
\lambda_{\Phi_e,l}-\lambda_{\Phi_g,l} = f(\epsilon)\left[ \lambda_{e,l}-\lambda_{g,l} \right] + g(\epsilon)\lambda_{e,g,l},
\end{equation}
where
\begin{equation}
f(\epsilon) = 1- 2\epsilon^2 , \; \hspace{1 cm} \; g(\epsilon) = 4\epsilon \sqrt{1-\epsilon^2} \cos \theta.
\end{equation}
The coupling constants $\lambda_{g,l}$, $\lambda_{e,l}$, and $\lambda_{e g,l}$ are the electron-phonon coupling constants associated to the unperturbed electronic states 
$|\Psi_{g}^{(0)} \rangle$ and $|\Psi_{e}^{(0)} \rangle$, respectively. Here $\theta$ is an arbitrary phase and $\epsilon$ is a mixing parameter approximately given by
\begin{equation}
\epsilon \approx {\langle e | V_u | g \rangle \over \hbar\left(\omega_{eg} - \omega_{\mbox{\scriptsize ph}} \right)},
\end{equation}
where $V_u$ is the intensity of the periodic perturbation perturbation, $\hbar \omega_{eg}$ is the electronic gap between the excited and ground states, and $\hbar\omega_{\mbox{\tiny ph}}$ is the energy of the phonon mode. For the SiV$^-$ center $\hbar \omega_{eg} = 1.68$ eV and $V_u \ll \hbar \omega_{eg}$, therefore we expect that $|\epsilon | \ll 1$. By symmetry considerations only phonons with character odd or even contribute to the effective coupling constants $\lambda_{e,l}-\lambda_{g,l} $ or $ \lambda_{e,g,l}$, respectively. As a consequence of both symmetry constraints we deduce that $\left( \lambda_{e,l}-\lambda_{g,l} \right) \lambda_{e,g,l} = 0$ for each lattice mode $l$. Finally, taking the limit $|\epsilon|\ll 1$ and averaging over the phase the spectral density function is
\begin{equation}
J(\omega) = \sum_{l} \left(\lambda_{\phi_e,l}-\lambda_{\phi_g,l} \right)^2\delta(\omega - \omega_l) = J_0(\omega) + 8 \epsilon^2  J_{eg}(\omega)
\end{equation}
and we recover the spectral density function given in Eq.\eqref{Totalspectral density function}.

\bibliographystyle{unsrt}

\end{document}